\documentclass[9pt,twocolumn,twoside]{osajnl}
\journal{ol} 
\setboolean{shortarticle}{true}
\title{Depth-Resolved Speckle-Correlations Imaging through Scattering Layers}

\author[1$^\dagger$]{Ofer Salhov}
\author[1$^\dagger$]{Gil Weinberg}
\author[1,*]{Ori Katz}

\affil[1]{Department of Applied Physics, Hebrew University of Jerusalem, Jerusalem 9190401, Israel}

\affil[*]{Corresponding author: orik@mail.huji.ac.il}
\affil[$^\dagger$]{These authors contributed equally to this work}

\ociscodes{(290.0290) Scattering; (030.6140) Speckle}

\begin{abstract}
Recently, novel imaging techniques based on the 'memory-effect' speckle-correlations have enabled diffraction-limited imaging through scattering layers and around corners. These techniques, however, are currently limited to imaging only small planar objects that are contained within the angular and axial range of the memory effect. In addition, they do not provide depth information or depth sectioning capability.
Here, we extend speckle-correlation imaging to include high-resolution depth-sectioning capability in reflection-mode, by combining it with coherence-gating via low coherence holography. We demonstrate depth measurements of hidden targets through a scattering layer, and speckle-correlation imaging using coherence-gated scattered light.
\end{abstract}

\setboolean{displaycopyright}{true}

\begin{document}

\maketitle
Imaging objects hidden behind diffusive barriers or outside the line of sight is an important challenge with applications ranging from biomedical imaging to defense. Naturally, this challenge has been at the focus of many works in the field of optics the last decades.  
Approaches that rely on measuring only the unscattered 'ballistic' photons, such as optical coherence tomography (OCT) \cite{huang1991optical,jeong2018focusing,badon2016smart,woo2016depth}, confocal microscopy, or LiDAR, allow diffraction-limited imaging through moderately scattering media, only when a sufficiently high level of ballistic photons exist.
Techniques that utilize scattered light for imaging, such as photoacoustic tomography, acousto-optic tomography, diffuse optical tomography (DOT) \cite{ntziachristos2010going}, and time-of-flight based inverse problem solutions\cite{velten2012recovering,o2018confocal,gariepy2016detection}, allow imaging through visually-opaque scattering layers and around corners, but with a spatial resolution that is orders of magnitude lower than the optical diffraction-limit. Wavefront-shaping approaches that aim at actively correcting or 'inverting' the effects of scattering \cite{vellekoop2007focusing,mosk2012controlling} require a guide-star \cite{horstmeyer2015guidestar} and wavefront-modulators that are faster than the speckle decorrelation time, for effective focusing. 

Recently, novel approaches that rely on angular speckle-correlations known as the 'memory-effect'  \cite{feng1988correlations} have allowed diffraction-limited computational imaging of objects hidden behind scattering layers or around corners \cite{katz2014non,bertolotti2012non,edrei2016optical}. These technique make use of simple imaging setups to measure the scattered light patterns, and retrieve the high-resolution imaging information by analyzing correlations in these patterns. 
Surprisingly, in some scenarios, even a single image of multiply-scattered diffused light is sufficient for diffraction limited imaging \cite{katz2014non}.  

However, since these techniques rely on speckle correlations that have inherently a narrow angular and axial range, they are limited to imaging small, isolated, and nearly-planar objects that are contained within the narrow field of view (FoV) of the memory effect. For diffusive samples, the angular FoV is dictated by the ratio between the optical wavelength, $\lambda$, and the scattering layer thickness, $L$ : $FoV \approx \lambda/\pi L$, and the axial extent of the correlations is: $ \delta z=(2\lambda/\pi)(z/D)^2$, where $z$ is the distance to the imaged object from the scattering layer, and $D$ is the diameter on the diffusive surface that is used for light collection.
While some depth information can be obtained in some scenarios by utilizing multiple views \cite{takasaki2014phase,shi2017non}, these approaches still require the hidden objects to be all located within the memory-effect range.
The inability to image extended three-dimensional scenes severely limits the application of speckle-correlation imaging in many real-world scenarios.

\begin{figure*}[!h]
\centering
\includegraphics[width=0.68\textwidth,]{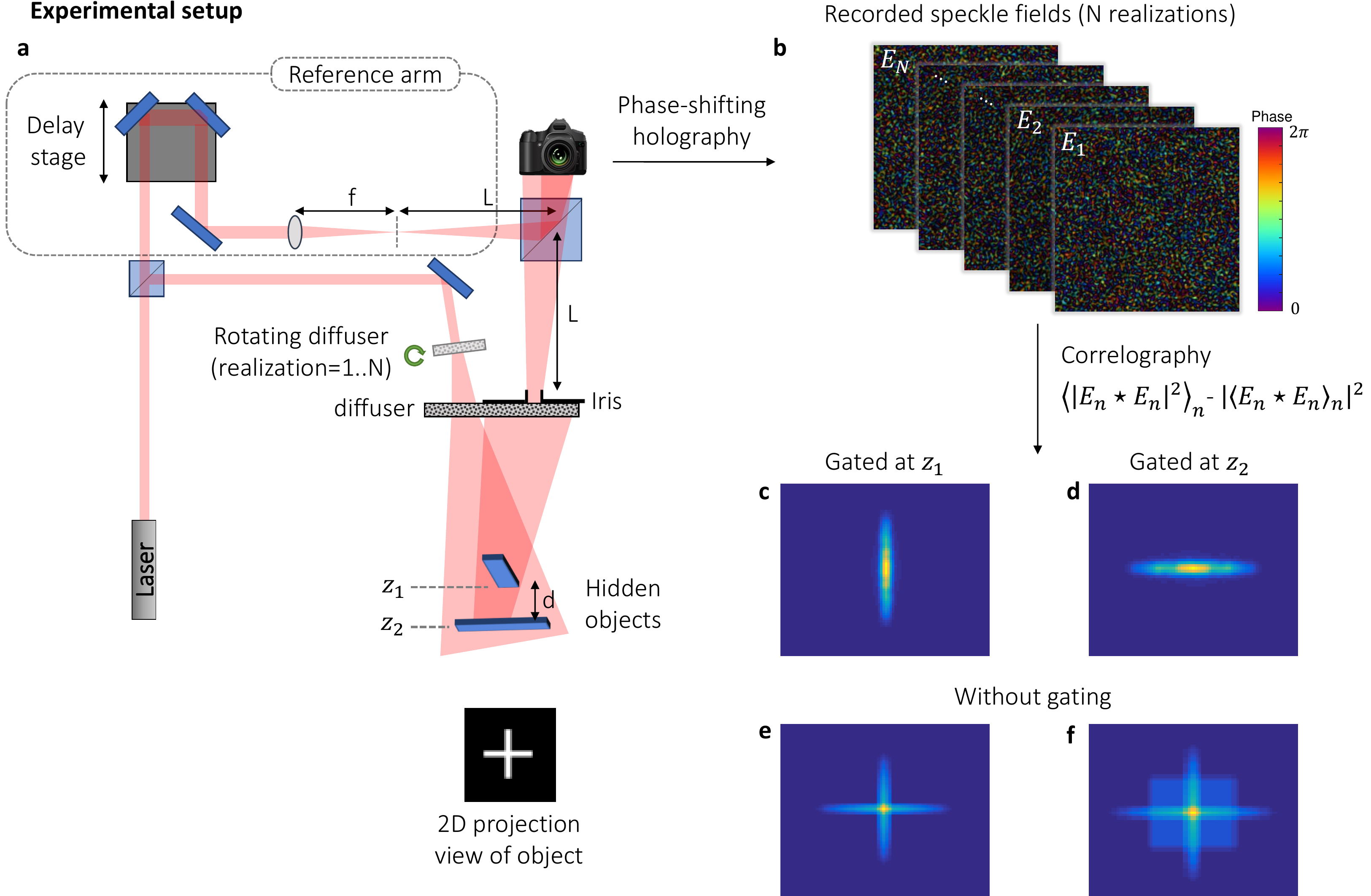}
\caption{\label{fig:setup}\textbf{Principle and numerical example of the proposed method. a,} Experimental setup: Two hidden objects are located at depths $z_1$ and $z_2=z_1+d$ behind a diffusive barrier (optical diffuser). A spatially-coherent, low temporal-coherence laser beam is split to two arms. In one arm, the beam illuminates the hidden objects through the diffuser, after passing through an additional computer-controlled rotating diffuser. The reflected light from the targets passes again through the diffusive barrier and is recorded by a camera. 
Depth sectioning is achieved by low-coherence holography: the scattered light on the camera is time-gated by interference with a beam from a reference arm, whose length is controlled to a selected target depth. At each depth, $N$ coherence-gated speckle fields are recorded at $N$ different realizations of the speckle illumination beam.
\textbf{b}, Simulated recorded coherence-gated fields. 
\textbf{(c-f)} Comparison of the autocorrelations calculated from the recorded fields via correlography \cite{idell1987image,edrei2016optical} with the proposed coherence-gating (c-d), and  without coherence-gating (e-f).
\textbf{(c,d)}, The coherence-gated autocorrelations provide depth-sectioning and retrieval of the hidden objects autocorrelations. 
\textbf{(e,f)}, Without time-gating, no sectioning or depth information is retrieved: 
 \textbf{e,} For $d$ larger than the speckle axial decorrelation length ($\delta z$) the autocorrelation is the sum of the objects' autocorrelations.  
\textbf{f}, For $d<\delta z$ the autocorrelation of the objects' 2D projection is obtained.}
\end{figure*}

Here, we extend speckle correlation imaging to allow depth-sectioning capability. We achieve this by performing speckle correlations analysis on coherence-gated scattered light. High resolution depth-sectioning is obtained in a similar manner to OCT: by recording the light fields that originate from a selected optical path difference (i.e. depth) behind a scattering layer \cite{woo2016depth}. However, unlike the coherence-gated images in conventional OCT, the coherence-gated fields in our experiments are random speckle patterns (Fig.1b) that are measured through highly scattering samples, and do not directly reveal the hidden objects. The hidden objects images at each selected depth are retrieved by speckle autocorrelation analysis \cite{katz2014non,edrei2016optical}. This combination of coherence-gating and speckle-correlation imaging allows us to eliminate the requirement for planar targets in speckle correlation imaging \cite{katz2014non}, or the requirement for measuring ballistic photons in OCT. 

In conventional speckle-correlation imaging \cite{katz2014non} a spatially-incoherent target object, located behind a thin scattering layer, is imaged by a camera located on the other side of the scattering layer. For objects with dimensions smaller than the memory effect range the scattering layer yields an effectively shift-invariant speckled point-spread function (PSF) $S(\theta)$, where $\theta$ is the viewing angle. The camera image is then given by: $I(\theta)=O(\theta)*S(\theta)$, where $I(\theta)$ is the camera image, $O(\theta)$ is the object intensity pattern, and $*$ denotes a convolution.  Since the autocorrelation of a speckle pattern, $S\star S $, is a sharply peaked function, taking the autocorrelation of the camera image $I\star I $ provides an estimate of the object  autocorrelation $O\star O $:
\begin{equation}\label{single_obj_autocorr}
[I\star I](\theta)\approx [O \star O](\theta) + a
\end{equation}
where $a$ is a constant background term. The object pattern is retrieved from its autocorrelation via phase-retrieval \cite{katz2014non,fienup1978reconstruction}.

The reason that conventional speckle correlation imaging cannot allow imaging of large, or multiple objects located at various depths, can be understood by considering the simple case of imaging two objects $O_1(\theta)$ and $O_2(\theta)$, that are separated in transverse or axial dimensions by a distance larger than the memory effect angular or axial range. In such a case, each object would be scattered by a different, uncorrelated PSF, $S_1(\theta)$ and $S_2(\theta)$, such that $[S_1 \star S_2](\theta)=b$. The resulting camera intensity distribution in this case would be: $I=O_1*S_1+O_2*S_2$ , yielding an autocorrelation that is the sum of the objects' autocorrelations, with no simple way to unmix them:  
    \begin{equation}\label{two_objects_autocorr}
    [I\star I](\theta) \approx [O_1 \star O_1](\theta) + [O_2 \star O_2](\theta) + c
    \end{equation}
where c is a constant background term, and we have used  $S_1 \star S_1=S_2 \star S_2 \propto \delta (\theta) + b $.
In such a scenario, the camera image autocorrelation cannot provide any information on the transverse or axial distance between the two objects, and it is not possible to reconstruct any of the hidden objects since their autocorrelations are mixed. 

In order to overcome this fundamental FoV limitation and in addition obtain depth-sectioning capability to speckle correlation imaging, we have developed an approach that combines speckle correlation imaging \cite{katz2014non,edrei2016optical} with temporal (coherence) gating via low-coherence holography \cite{woo2016depth}. In our approach, depth sectioning is obtained by selectively recording only speckles that originate from a selected optical path (depth). Angular speckle correlations of these coherence-gated fields are then analyzed in a manner similar to the speckle correlography approach of Edrei et al. \cite{edrei2016optical,idell1987image}. The correlography approach allows the retrieval of the same 'incoherent' intensity-autocorrelation as is obtained in the approach of \cite{katz2014non}, but using spatially-coherent, rather than incoherent, illumination, and with an increased SNR (speckle ensemble averaging) \cite{idell1987image}.

The setup used to implement the proposed approach is the phase-shifting low-coherence holography setup described in Fig. \ref{fig:setup}a. The illumination is provided by a laser diode (Toptica iBeam-SMART-640-S G1) having a single transverse mode at a central wavelength of 640nm, and a coherence length of $l_{coh} \approx 0.4mm$. 
The laser beam is split into two arms of a Michelson interferometer: the first arm is a conventional speckle-correlation imaging arrangement in reflection mode, with the addition of a rotating diffuser (10DKIT-C1 1° and 0.5°, Newport) mounted on a controlled rotation stage before the diffusive barrier (optical diffuser, 10DKIT-C1 5°, Newport), to provide different speckle realizations of the illumination beam \cite{idell1987image,edrei2016optical}. The scattered light illuminates the hidden scene made of reflective diffusive objects placed at different depths. The reflected light from the objects travels again through the diffusive barrier and is recorded by a camera placed at a distance $v=115mm$ from the barrier.  

The second arm of the interferometer is a reference arm containing a computer controlled delay-stage, which is used to select the target depth, and to perform phase-shifting holography. As in OCT, demodulating the low-coherence interference as a function of the delay between the two arms, allows selective measurement of only the scattered light arriving at a specific chosen time delay, i.e. depth, with a depth resolution given by the laser coherence length.
The reference arm also contains a focusing lens, which focuses the reference beam at a distance $v$ from the camera plane, identical to the camera-diffuser distance. This ensures that the reference beam has the same curvature as light scattered from a point on the diffuser, allowing to recover the field at the diffuser plane by a simple Fourier transform of the measured fields.  

The principle of our approach is presented in Fig.1(b-d), and proof of principle experimental results are presented in Figures 2-3.
To achieve coherence-gated speckle-correlation images the following steps are taken:
(1) The reference arm delay is set to the target depth. 
(2) The rotating diffuser position is set to provide $N\gg1$ different speckle illumination patterns (realizations). For each realization, $i$, the coherence-gated field $E_{i}$ (Fig.1b) is acquired by phase-shifting holography\cite{yamaguchi1997phase} from $M \geq 3$ camera images, $I_m$, where $m=1..M$: 
\begin{equation}\label{phase_shifting}
E_m=\frac{\frac{1}{M} \sum_{m=1}^M I_me^{i 2\pi m/M}}  {\sqrt[]{I_{ref}}} 
  \end{equation}where $I_{ref}$ is the intensity pattern of the reference beam, measured by blocking the other arm. (3) Since the 2D Fourier-transform of $E_i$ provides the field at the diffuser plane, the intensity-autocorrelation of the hidden object $O(\theta)$ at the chosen target depth, can be retrieved from $E_i$ in the same manner as was demonstrated by Edrei et al.\cite{edrei2016optical}, using correlography \cite{idell1987image}: 
\begin{equation}\label{correlography}
O\star O \approx \left\langle \left|E_i\star E_{i}\right|^{2}\right\rangle _{i}-\left|\left\langle E_{i}\star E_i\right\rangle _{i}\right|^{2} 
\end{equation}
 where the first term in \ref{correlography} provides the (coherent) field-autocorrelations of the object, and the second term subtracts the coherent autocorrelation peak (the autocorrelation of a single speckle grain).
(4) Finally, the target at each chosen depth can be reconstructed from its autocorrelation (Fig.1c,d) via phase-retrieval \cite{fienup1978reconstruction}.
 
Our imaging approach is thus identical to the correlography approach of Edrei et al. \cite{edrei2016optical}, with the important fundamental difference that is the addition of the low-coherence based temporal-gating. Thus, images reconstructed at different time-gates in our approach (Fig.1c-d) carry additional 3D information compared to the reconstructions possible by simple correlography (Fig.1e-f). 

\begin{figure}[!h]
\centering
\includegraphics[width=0.4\textwidth]{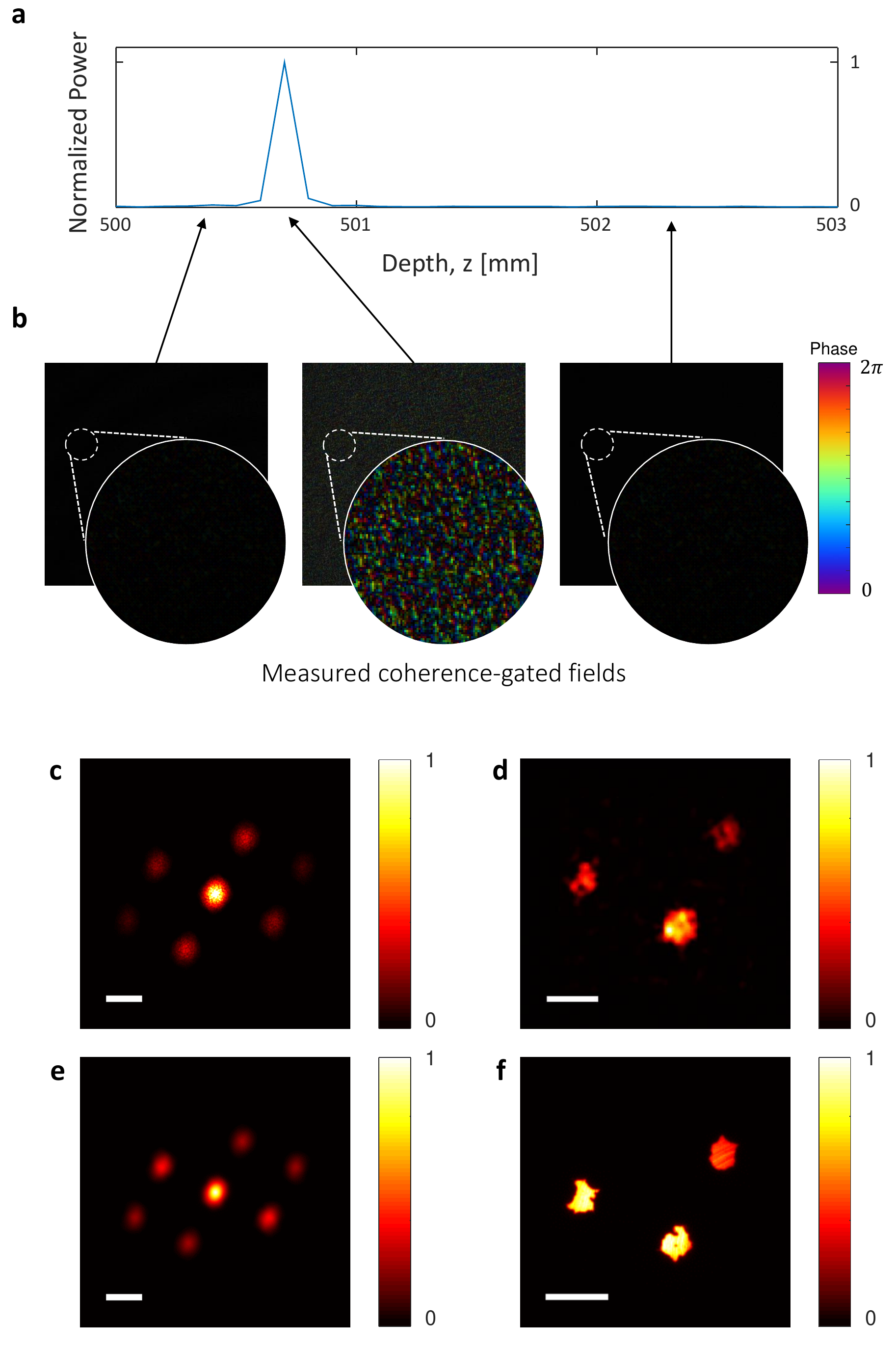}
\caption{\label{fig:scan}\textbf{Experimental depth-resolved imaging of a single hidden object. a}, Total intensity of the time-gated fields as a function of depth (distance from diffuser) reveals a single target at $z=500.7mm$ depth. \textbf{b}, Recorded time-gated fields at several depths. \textbf{c}, Intensity autocorrelation calculated from $N=40$ fields at $z=500.7mm$. \textbf{d}, Phase-retrieval reconstruction from the autocorrelation of (c); \textbf{e}, hidden object autocorrelation; \textbf{f}, hidden object. Scale bars: 1.25mm.}
\end{figure}

Figures 2-3 present two proof-of-principle experiments of the proposed approach, with either a single hidden planar object (Fig.2) or two planar objects placed at two different depths (Fig.3). In order to initially locate the targets depths in each experiment, a delay scan is performed using a single illumination speckle realization: for each delay, the coherence-gated field is measured by phase-shifting interferometry (Fig.\ref{fig:scan}b), and its total intensity is plotted as a function of delay.  The peaks in the plotted trace reveal the target's depth (Fig.2a). The delay is then set to the selected target depth, and the target at the chosen delay (Fig.2d)  is reconstructed from its intensity-autocorrelation (Fig.2c), calculated from $N=40$ coherence gated fields measured at the targeted depth for $N$ different realizations of the illumination. 

The results presented in Figure 3 demonstrate the depth-sectioning capability of the proposed approach in the case of multiple targets placed at different depths. Our approach allows selective retrieval of the autocorrelation of each individual target at the depth selected via coherence-gating. The retrieved autocorrelation (Fig.3e,g) reveals only the autocorrelation of the target at the chosen depth (Fig.3f,h), which would be otherwise mixed with the autocorrelations of targets located at different depth, as is the case when using conventional speckle-correlation imaging (Fig.3c,d)\cite{edrei2016optical,katz2014non}. In (e-g) the residual coherence peak was reduced by an optimal threshold \cite{edrei2016optical}. 

\begin{figure}[!h]
\centering
\includegraphics[width=0.4\textwidth]{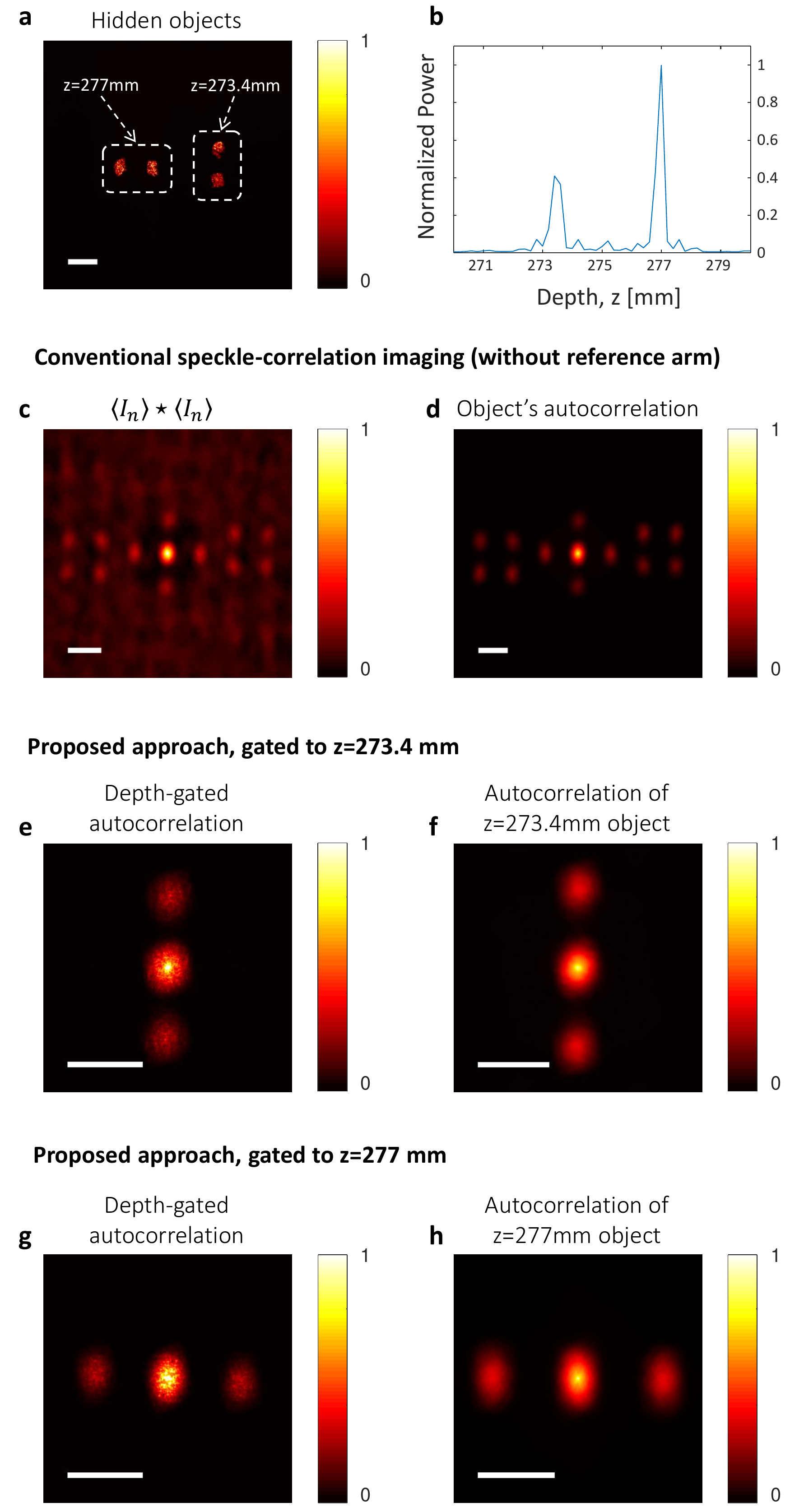}	
\caption{\label{fig:retreival}\textbf{Depth-sectioned speckle-correlation imaging. a}, Direct image of reflecting targets hidden at two depths, without the scattering layer. The depths obtained by OCT are marked. \textbf{b}, Total intensity of the coherence-gated fields as a function of depth reveals the two objects at $z_1=273.4mm$ and $z_2=277mm$.
\textbf{c}, Conventional speckle-correlation imaging \cite{katz2014non}  (without reference arm), using $N=60$ speckle realizations reveals the object's autocorrelation (d) but does not provide depth information or sectioning. \textbf{d}, Autocorrelation of (a). \textbf{e}, Autocorrelation obtained with the proposed approach coherence-gated to $z=273.4mm$ depth, revealing the  autocorrelation of only the objects located at the selected depth \textbf{(f)}. \textbf{g-h}, same as (e-f), for target depth $z=277mm$. Scale bars: (a) 1.8mm, (c-h) 120 camera pixels.}
\end{figure} 

We have presented an approach for adding depth-sectioning capability to speckle-correlation imaging. 
Beyond allowing depth retrieval and sectioning, our approach relaxes the requirement for the object's axial dimension to be smaller than the speckle axial correlation length $\delta z$  \cite{katz2014non}. 
For our approach to effectively work, it is required that $\delta z > l_c$, where $l_c $ is the coherence length of the light source (the depth-sectioning resolution). Another requirement is that the object's angular dimensions at each depth, rather than at \textit{all} depths together, should be smaller than the memory-effect angular range (FoV). As in \cite{katz2014non,idell1987image}, for accurate estimation of the object autocorrelation, the number of speckles captured on the camera should be maximized.
The number of speckle grains is limited by the number of pixels on the camera sensor. In order to maximize it we have used a high resolution camera and a phase-shifting holography approach (rather than off-axis holography). 
Utilizing the $N$ different speckle realizations in the correlography approach yields a further improvement of the speckle autocorrelation fidelity by a factor of $\sqrt[]{N}$ on its statistical signal-to-noise (ensemble averaging)  \cite{idell1987image}.
Additionally, in order to ensure that the conditions for correlography are met, the diffusive medium must be located at a large enough distance from the targets. Using speckle illumination, this 'far-field' condition is $z>2D\cdot r_c/\lambda$ , where $r_c$  is the correlation radius of the illumination speckle pattern at the object plane (speckle grain size), and $D$ is  the transverse dimension of the object.
A notable limitation of this method is its inability to distinguish between transverse translations of a  reconstructed section, as this information is ignored by the autocorrelation operation, making full 3D reconstruction challenging without additional information. The iterative phase-retrieval algorithm can be replaced by a deterministic reconstruction by performing bi-spectrum based analysis \cite{wu2016single}.

We used coherence-gating to enable depth-sectioning by recording time-gated speckle patterns. However, other temporal-gating approaches, such as LiDAR \cite{gariepy2016detection,velten2012recovering,o2018confocal} may be used as well for this goal. 

This work was supported  by the DARPA REVEAL program under contract number HR0011-16-C-0027. The information presented in this paper does not necessarily reflect the position or policy of either DARPA or the US Government, and no such official endorsement should be inferred.

\bibliography{references}

\bibliographyfullrefs{references}

\end{document}